\documentclass{aa}
\usepackage{graphicx}
\usepackage{natbib}
\usepackage{amssymb}
\usepackage{amsfonts}
\usepackage{amsbsy}
\usepackage{amsmath}

\bibpunct{(}{)}{;}{a}{}{,}

\newcommand{\PFA}{P_{\rm FA}}
\newcommand{\PD}{P_{\rm D}}

\newcommand{\oneb}{\boldsymbol{1}}
\newcommand{\ab}{\boldsymbol{a}}

\newcommand{\Bb}{\boldsymbol{B}}

\newcommand{\Fb}{\boldsymbol{F}}
\newcommand{\Gb}{\boldsymbol{G}}

\newcommand{\eb}{\boldsymbol{e}}
\newcommand{\gb}{\boldsymbol{g}}
\newcommand{\Cb}{\boldsymbol{C}}

\newcommand{\Db}{\boldsymbol{D}}
\newcommand{\Hb}{\boldsymbol{H}}
\newcommand{\Ib}{\boldsymbol{I}}

\newcommand{\Jb}{\boldsymbol{J}}

\newcommand{\nb}{\boldsymbol{n}}

\newcommand{\Deltab}{\boldsymbol{\Delta}}
\newcommand{\Sigmab}{\boldsymbol{\Sigma}}

\newcommand{\Pb}{\boldsymbol{P}}

\newcommand{\ssb}{\boldsymbol{s}}

\newcommand{\Sb}{\boldsymbol{S}}

\newcommand{\xb}{\boldsymbol{x}}

\newcommand{\yb}{\boldsymbol{y}}
\newcommand{\Tb}{\boldsymbol{T}}
\newcommand{\Zb}{\boldsymbol{Z}}
\newcommand{\ub}{\boldsymbol{u}}
\newcommand{\Ub}{\boldsymbol{U}}
\newcommand{\wb}{\boldsymbol{w}}

\newcommand{\zerob}{\boldsymbol{0}}
\newcommand{\ooneb}{\boldsymbol{1}}
\newcommand{\lambdab}{\boldsymbol{\lambda}}
\newcommand{\Lambdab}{\boldsymbol{\Lambda}}

\newcommand{\Chb}{\boldsymbol{\widehat C}}
\newcommand{\xhb}{\boldsymbol{\widehat x}}
\newcommand{\shb}{\boldsymbol{\widehat s}}

\newcommand{\xtb}{\boldsymbol{\widetilde x}}

\newcommand{\ztb}{\boldsymbol{\widetilde z}}
\newcommand{\stb}{\boldsymbol{\widetilde s}}
\newcommand{\Gtb}{\boldsymbol{\widetilde G}}

\newcommand{\Ztb}{\boldsymbol{\widetilde Z}}
\newcommand{\Sigmatb}{\boldsymbol{\widetilde \Sigma}}

\newcommand{\Hc}{\mathcal{H}}

\newcommand{\Nmatc}{\mathcal{N}}

\newcommand{\xbl}{\boldsymbol{\underline{x}}}
\newcommand{\gbl}{\boldsymbol{\underline{g}}}
\newcommand{\Gbl}{\boldsymbol{\underline{G}}}
\newcommand{\gl}{\underline{g}}

\newcommand{\ssbl}{\boldsymbol{\underline{s}}}
\newcommand{\nbl}{\boldsymbol{\underline{n}}}
\newcommand{\ubl}{\boldsymbol{\underline{u}}}
\newcommand{\rtbl}{\boldsymbol{\underline{\widetilde r}}}
\newcommand{\stbl}{\boldsymbol{\underline{\widetilde s}}}
\newcommand{\gtbl}{\boldsymbol{\underline{\widetilde g}}}
\newcommand{\Gtbl}{\boldsymbol{\underline{\widetilde G}}}

\newcommand{\utbl}{\boldsymbol{\underline{\widetilde u}}}
\newcommand{\xtbl}{\boldsymbol{\underline{\widetilde x}}}
\newcommand{\Utbl}{\boldsymbol{\underline{\widetilde U}}}

\newcommand{\rbl}{\boldsymbol{\underline{r}}}
\newcommand{\Cbl}{\boldsymbol{\underline{C}}}
\newcommand{\Sigmatbl}{\boldsymbol{\underline{\widetilde \Sigma}}}

\newcommand{\Smatb}{\boldsymbol{{\mathcal S}}}
\newcommand{\Xmatb}{\boldsymbol{{\mathcal X}}}
\newcommand{\Ymatb}{\boldsymbol{{\mathcal Y}}}
\newcommand{\Bmatb}{\boldsymbol{{\mathcal B}}}
\newcommand{\Cmatb}{\boldsymbol{{\mathcal C}}}
\newcommand{\Ematb}{\boldsymbol{{\mathcal E}}}
\newcommand{\Fmatfb}{\boldsymbol{{\mathfrak F}}}
\newcommand{\Fmatb}{\boldsymbol{{\mathcal F}}}
\newcommand{\Gmatb}{\boldsymbol{{\mathcal G}}}
\newcommand{\Nmatb}{\boldsymbol{{\mathcal N}}}
\newcommand{\Umatb}{\boldsymbol{{\mathcal U}}}

\begin{document}

   \title{A theoretical framework for the detection of \\ point-sources in Cosmic Microwave Background maps}

   \author{R. Vio\inst{1}
    \and
           P. Andreani\inst{2}
          }
   \institute{Chip Computers Consulting s.r.l., Viale Don L.~Sturzo 82,
              S.Liberale di Marcon, 30020 Venice, Italy\\
              \email{robertovio@tin.it},
         \and
		  ESO, Karl Schwarzschild strasse 2, 85748 Garching, Germany\\
                  INAF-Osservatorio Astronomico di Trieste, via Tiepolo 11, 34143 Trieste, Italy\\              
		  \email{pandrean@eso.org}
             }

\date{Received .............; accepted ................}

\abstract
{}
{The detection of point-sources in experimental microwave maps is a critical step in the analysis of the Cosmic Microwave Background (CMB) data. If not properly
removed, these sources can have adverse effects on the estimation of the power-spectrum and/or the test of Gaussianity of the CMB
component. In the literature, various techniques have been presented to extract point sources from an observed image but no 
general consensus about their real performance and properties has been reached. Their characteristics have been 
studied essentially through numerical simulations based on semi-empirical models of the CMB and the Galactic foreground. Such models often have 
different levels of sophistication and/or are based on different physical assumptions (e.g. the number of Galactic components and 
level of the noise). Moreover, the application of a given technique to a set of data (either simulated or experimental) requires
the tuning of one or more parameters that unavoidably is a subjective operation. Hence, a reliable comparison is difficult. 
What is missing is a statistical analysis of the properties of the proposed methodologies.
This is the aim of the present paper.}{The statistical properties of the detection techniques in the context of two different 
criteria, i.e. the Neyman-Pearson
criterion and the maximization of the signal-to-noise ratio (SNR), are analyzed through an analytical 
approach. One-dimensional as well two-dimensional signals are considered. The case of multiple observing frequencies
is also addressed.}{The conditions are fixed under which the techniques can work satisfactorily. Their limits are also illustrated
and implementation details provided. We show that, exploiting some a priori information, it is possible
to develop simple algorithms with performances not too far from those of more sophisticated but complex techniques. 
In this respect, a detection algorithm, tailored for high Galactic latitudes, 
is presented that could be useful in future 
ground-based experiments as, for example, the Atacama Large Millimeter/submillimeter Array (ALMA).}
{}
\titlerunning{Point-Source Detection}
\authorrunning{R. Vio, \& P. Andreani}
\maketitle

\section{Introduction}

The detection of point-sources embedded in a noise background is a critical issue
in the analysis of the experimental Cosmic Microwave Background (CMB) maps. The estimation of the power-spectrum of the CMB component and the test of its possible 
nonGaussian nature need the a priori detection and removal of these sources. In particular the former operation is rather delicate.
Given its importance, this subject has been extensively considered in literature
\citep[see][ and references therein]{her08a}. However, no widespread accepted conclusion has been reached.
The reason is that most of the detection techniques presented in literature lack a sufficiently
rigorous theoretical background. The statistical characteristics are derived from numerical experiments only. 
Especially
in the experiments where it is necessary to simulate maps at different observing frequencies, this approach is
hazardous for two reasons: 1) the models used by the various authors in the simulations are not the same; 2) when a detection technique has to be
applied to a set of data (either synthetic or real) the tuning of some parameters is unavoidably subjective. Because of this contrasting results
appear in literature. A safer procedure consists in studying the techniques in a well defined theoretical framework. 
Although some of the conditions assumed to fix the theoretical context could be not realistic, anyhow it is possible to obtain a more 
objective comparison
as well as some useful indications that help to understand what it is possible to expect when they are relaxed. The aim of
the present paper is to provide a theoretical characterization of some of the detection techniques suited for both single and 
multiple-frequency CMB experiments.
For ease of notation, initially the arguments will be developed for one-dimensional signals
$\xb = ( x[0], x[1], \ldots, x[N-1])^T$. Later they will extended to the two-dimensional situation.
The paper is organized as follows: in Sec.~\ref{sec:formalization} the problem of point-source detection is described in detail
for the case of a single spatial dimension. Here some standard material is presented with the aim to fix notation and formalism. 
In Sec.~\ref{sec:multiple} the situation is considered when more signals are available. 
In Sec.~\ref{sec:CMB} a technique tailored for observation at high Galactic latitude is presented that will be useful
in some experiments planned in the near future with innovative instruments as the Atacama Large Millimeter/submillimeter Array  (ALMA). 
Finally, the conclusions are given in Sec.~\ref{sec:conclusions}. In appendix~\ref{sec:efficient}, two procedures are
presented that efficiently implement the techniques described in the text. These are
extended to the two-dimensional case in appendix~\ref{sec:twodimensional}.

\section{Formalization of the problem} \label{sec:formalization}

The first step in the development of a detection technique is to fix the conditions that are pertinent to the problem
of interest. In the case of CMB observations, the following conditions are commonly assumed:
\begin{enumerate}
\item The point-sources have a known spatial profile $\ssb = a \gb$. The amplitude ``$a$'' is a scalar quantity 
different from source to source, whereas $\gb$ is a function, due to the instrument beam,
that is identical for all of them. Function $\gb$ is normalized in such a way that $\max{ \{ g[0], g[1], \ldots, g[N-1] \} } = 1$; \\
\item The point-sources are embedded in a noise-background, i.e. the observed signal $\xb$ is given by $\xb = \ssb + \nb$. In other words, 
noise is additive. Usually, this is a reasonable assumption; \\
\item Noise $\nb$ is the realization of a stationary stochastic process with known covariance matrix
\begin{equation} \label{eq:C}
\Cb = {\rm E}[\nb \nb^T]. 
\end{equation}
Actually, because of the Galactic contribution, especially
at low Galactic latitudes, this hypothesis is not satisfied. However, it is assumed to hold locally. This allows 
the computation of statistics as the mean or the covariance matrix that, otherwise, should not be possible. 
Without loss of generality, it is assumed that ${\rm E}[\nb] = 0$.
\end{enumerate}   
Under these conditions, the detection problem consists in deciding whether $\xb$ is a pure
noise $\nb$ (hypothesis $H_0$) or it contains also the contribution of a source $\ssb$ 
(hypothesis $H_1$). In other words, the source detection problem is 
equivalent to a decision problem where two hypotheses hold:
\begin{equation} \label{eq:decision}
\left\{
\begin{array}{ll}
\Hc_0: & \quad \xb = \nb; \\
\Hc_1: & \quad \xb = \nb + \ssb.
\end{array}
\right.
\end{equation}
Under $\Hc_0$ the probability density function of $\xb$ is given by $p(\xb| \Hc_0)$ whereas under $\Hc_1$ by $p(\xb| \Hc_1)$.
At this point, it is necessary to fix the criterion to use for the detection. Clearly, one cannot hope to find all the sources present in a given 
signal. Hence, some choices are necessary. For example, one could decide that the non-detection or the misidentification of a bright source could be
more important than that of a fainter one, or vice versa. A very common and effective criterion is the Neyman-Pearson criterion that consists in the
maximization of the {\it probability of detection} $\PD$ 
under the constraint that the {\it probability of false alarm} $\PFA$ (i.e., the probability of a false detection) does not exceed a fixed value
$\alpha$. The Neyman-Pearson theorem \citep[e.g., see ][]{kay98} is a powerful tool that allows to design
a decision process that pursues this aim:
{\it to maximize $\PD$ for a given $\PFA=\alpha$, decide $\Hc_1$ 
if the {\it likelihood ratio} (LR)
\begin{equation} \label{eq:ratio}
L(\xb) = \frac{p(\xb| \Hc_1)}{p(\xb| \Hc_0)} > \gamma,
\end{equation}
where the threshold $\gamma$ is found from
\begin{equation} \label{eq:p1}
\PFA = \int_{\{\xb: L(\xb) > \gamma\}} p(\xb| \Hc_0) d\xb = \alpha.
\end{equation}
}
The test of the ratio~\eqref{eq:ratio} is called the {\it likelihood ratio test} (LRT). 

An important example of application of LRT is when noise $\nb$ is Gaussian with correlation function
$\Cb$. Actually, in CMB experiments this condition is satisfied only for observations at high Galactic latitudes where the CMB emission
and the instrumental noise are by far the dominant contributors. At lower latitudes, it is often assumed 
to hold locally. For example, the contribution to $\xb$ of components that in small patch of sky presents linear spatial 
trends are often approximated with stationary Gaussian processes with a steep spectrum (e.g. $1/f$ noises).
In any case, even if the Gaussianity condition was unrealistic, it is often made anyway since it allows an analytical treatment of the problem 
of interest and the results can be used as a benchmark in the analysis of more complex scenarios. With Gaussian $\nb$, it is
\begin{align}
p(\xb | \Hc_0) & = \Delta
\exp\left[ -\frac{1}{2} \xb^T \Cb^{-1} \xb \right]; \label {eq:t1} \\
p(\xb | \Hc_1) & = \Delta  
\exp\left[ -\frac{1}{2} (\xb - \ssb)^T \Cb^{-1} (\xb -\ssb) \right], \label{eq:t2}
\end{align}
with
\begin{equation}
\Delta = \frac{1}{(2 \pi)^{\frac{N}{2}} {\rm det}^{\frac{1}{2}}(\Cb)}.
\end{equation}
The LRT is given by
\begin{equation}
l(\xb) = \ln[L(\xb)] = \xb^T \Cb^{-1} \ssb - \frac{1}{2} \ssb^T \Cb^{-1} \ssb > \gamma. 
\end{equation}
Hence, it results that
$\Hc_1$ has to be chosen when for the statistic $T(\xb)$ (called {\it NP detector}) it is
\begin{equation} \label{eq:test}
T(\xb) = \xb^T \Cb^{-1} \ssb > \gamma,
\end{equation}
with $\gamma$ such as
\begin{equation} \label{eq:pfa}
\PFA = Q \left( \frac{\gamma}{\left[ \ssb^T \Cb^{-1} \ssb \right]^{1/2}} \right) = \alpha,
\end{equation}
i.e.,
\begin{equation} \label{eq:gammap}
\gamma = Q^{-1}(\PFA) \sqrt{ \ssb^T \Cb^{-1} \ssb }.
\end{equation}
Here, $Q(x) = 1 - \Phi(x)$ with $\Phi(x)$ the standard Gaussian {\it distribution function}, $Q^{-1}(.)$ is
the corresponding inverse function. Equation~\eqref{eq:pfa} is due to the fact that $T(\xb)$ is a Gaussian random variable with 
variance $\ssb^T \Cb^{-1} \ssb$ and expected values equal to zero under $\Hc_0$ and $\ssb^T \Cb^{-1} \ssb$ under $\Hc_1$.
For the same reason it is
\begin{equation} \label{eq:pd}
\PD = Q \left( Q^{-1} \left( \PFA \right) - \sqrt{ \ssb^T \Cb^{-1} \ssb} \right).
\end{equation}
Equation~\eqref{eq:test} can be written in the form
\begin{equation} \label{eq:test1}
T(\xb) = \xb^T \ub > \gamma,
\end{equation}
with 
\begin{equation} \label{eq:mf}
\ub = \Cb^{-1} \ssb.
\end{equation}
From this equation appears that $\ub$ can be thought as a linear filter of signal $\xb$, that is called {\it matched filter} (MF).

\subsection{Some comments on the use of the matched filter in practical applications} \label{sec:comments}

There are some important points to stress with regard the MF when used in practical aplications. They are:
\begin{itemize}
\item $T_1(\xb)$ is a {\it sufficient statistic} \citep{kay98}.
Loosely speaking, this means that $T_1(\xb)$ is able to summarize all the relevant information in the data
concerning the decision~(\ref{eq:decision}). No other statistic can perform better.
As a consequence, the claim that some filters (e.g. the {\it Mexican hat wavelet}, the {\it scale-adaptive filters}, 
the {\it biparametric scale-adaptive filters...}) are superior to MF according to the Neyman-Pearson criterion \citep{bar03, lop05} 
is not correct. It is the result of the use of imprecise theoretical arguments \citep[e.g. see][]{vio04}; \\

\item It is worth noticing that if the amplitude ``$a$'' of the source is unknown,
then Eq.~(\ref{eq:test}) can be rewritten in the form
\begin{equation} \label{eq:test2}
T(\xb) = \xb^T \Cb^{-1} \gb > \gamma',
\end{equation}
with $\gamma' = \gamma / a = Q^{-1}(\PFA) \sqrt{ \gb^T \Cb^{-1} \gb }$. In other words, a statistic is  
obtained that is independent of ``$a$'', i.e.
also in the case that the amplitude of the source is unknown, $T(\xb)$  maximizes $\PD$
for a fixed $\PFA$. The only consequence is that $\PD$ cannot be evaluated in advance. In principle this
quantity could be evaluated a posteriori by using the maximum likelihood estimate of the amplitude, 
$\widehat{a} = \xb^T \Cb^{-1} \gb / \gb^T \Cb^{-1} \gb$, but this is of little interest.  
More useful is that in real experiments
one is typically interested in the detection of sources which have amplitudes characterized by a probability density function 
$p(a)$. In this case, once $\PFA$ is fixed to a
value $\alpha$ and making to change ``$a$'' across the domain of $p(a)$, the quantity $1-\PD$, with $\PD$ as given by 
Eq.~(\ref{eq:pd}) and $\ssb = a \gb$, provides an estimate of the fraction of undetected sources as function of their amplitude; \\

\item If $\shb = \Hb \ssb$ and $\xhb = \Hb \xb$, then
\begin{align}
T(\Hb \xb) & = \xhb^T \Chb^{-1} \shb \nonumber  \\
& = \xb^T \Hb^T \Hb^{-T} \Cb^{-1} \Hb^{-1} \Hb \ssb = T(\xb),
\end{align}
with $\Hb$ any invertible linear operator (matrix).
A useful consequence of this property is that if signal $\xb$ is convolved with a function (e.g., the beam of an instrument), 
this operation does not modify the optimality of MF. This fact could be useful in situations where more signals are available
that are obtained with different point spread functions (see below); \\

\item The arguments above are developed under the implicit assumption that sources do not overlap and that their position is known in advance. 
In practical applications the first condition is satisfied -- strictly speaking -- only in very high resolution maps and it is assumed to be
always valid for those sources above the confusion noise. The second statement, i.e. that the position of the sources is known in advance,
is not true and the standard procedure consists in filtering $\xb$ by means of $\ub$ and in
computing $T(\xb)$ for the peaks in the resulting signal. Although there is no guarantee that a peak in the filtered signal marks the true position
of a source even in the case this is effectively present, theoretical arguments as well as years of application in real-life problems 
have proved that this procedure is rather robust and able to provide excellent results; \\ 

\item If the Gaussianity of $\nb$ is relaxed, then MF is no longer optimal in the Neyman-Pearson sense. However, it
remains optimal with respect to the signal-to-noise ratio (SNR) \footnote{Here, the quantity {\rm SNR} is defined as the
ratio between the squared amplitude of the filtered source
with the variance of the filtered noise.}. This means that,  independently of the nature of the noise, MF provides the greatest amplification
of the signal with respect to the noise. This can be easily verified through the minimization of 
the variance of the filtered noise $\ub^T \nb$
with the constraint that $\ub^T \ssb = a$ (i.e. filter $\ub$ does not modifies the amplitude of the source), in formula
\footnote{We recall that the functions ``$\arg\min F(x)$'' and ``$\arg\max F(x)$''
provide the values of $x$ of for which the function $F(x)$ has the smallest and greatest value, respectively.}
\begin{equation}
\ub_{\rm SNR} = \underset{ \ub }{\arg\min}[\ub^T \Cb \ub - \lambda (\ssb^T \ub - a)] \label{eq:snr}
\end{equation}
with $\lambda$ a {\it Lagrange multiplier}. It is not difficult to see that
\begin{align}
\ub_{\rm SNR} & = a \Cb^{-1} \ssb / [\ssb^T \Cb^{-1} \ssb], \\
              & = \Cb^{-1} \gb / [\gb^T \Cb^{-1} \gb],
\end{align}
i.e. apart from a normalizing factor, $\ub_{\rm SNR}$ is given by Eq.~\eqref{eq:mf}. Since $\ub_{\rm SNR}$ is optimal as concerns the maximization
of the SNR, no other methods can outperform it in this respect. For this reason, expedients as the introduction of additional constraints 
and/or of free parameters in $\ub$
\citep[e.g. see][]{san01, her02} has the only effect to reduce the detection performances \citep[e.g. see][]{vio02}. 
One of the benefit in using $\ub_{\rm SNR}$ is that
the value of $\ub_{\rm SNR}^T \ssb$ provides directly an unbiased estimate of the amplitude ``$a$'' of the source. However, it is necessary
to stress that in practical applications, where the true position of the source is not known and it is necessary to apply the
procedure described above, this is no longer true; \\

\item When $\ssb$ is a long signal, some computational problems come out. In fact, if the size of $\ssb$ is $N$, then $\Cb$ is a 
$N \times N$ matrix. Hence, the computation
of the quantity $\xb^T \Cb^{-1} \ssb$ can become quite expensive. The computational burden can be alleviated if $\Cb$, that is a 
{\it Toeplitz} matrix, is approximated with a {\it circulant} matrix $\Cmatb$. This is because $\Cmatb$ can be diagonalized through
\begin{equation}
\Fb_{(N)} \Cmatb \Fb_{(N)}^H = \Sigmatb.
\end{equation}
Here, $\Fb_{(N)}$ is the $N \times N$ one-dimensional Fourier matrix that is a complex,
unitary, and symmetric matrix whose elements
are given by
\begin{equation}
(F_{(N)})_{kl} = \frac{1}{\sqrt{N}} {\rm e}^{-2 \pi \iota (k-1)(l-1) /N}.
\end{equation}
Symbol ``$\widetilde{~~}$'' denotes the one-dimensional Fourier transform, 
$\iota = \sqrt{-1}$, $\Fb_N^H$ is the complex conjugate transpose of $\Fb_N$, and $\Sigmatb$ is a diagonal matrix containing the eigenvalues
of $\Cmatb$ (i.e., the power-spectrum of $\nb$). After that, since $ \Fb_N$ is a unitary matrix, one obtains that
\begin{align}
\xb^T \Cb^{-1} \ssb & \approx (\xb^T \Fb_{(N)}^H) (\Fb_{(N)} \Cmatb \Fb_{(N)}^H)^{-1} (\Fb_{(N)} \ssb) = \label{eq:fft1} \\
& = \xtb^H \Sigmatb^{-1} \stb = \xtb^H ( {\rm DIAG}[\Sigmatb^{-1}] \odot \stb). \label{eq:fft2} 
\end{align}
Symbol ``$\odot$'' denotes the element-wise multiplication, and
${\rm DIAG}[\Zb]$ a column vector containing the diagonal elements of the square matrix $\Zb$.
In obtaining this result, the fact that $\Fb_{(N)} \Cmatb^{-1} \Fb_{(N)}^H = (\Fb_{(N)} \Cmatb \Fb_{(N)}^H)^{-1}$
has been used. In principle, all the terms in Eq.~\eqref{eq:fft2} could be computed quite efficiently by means of {\it fast Fourier transform} (FFT). 
In particular, array ${\rm DIAG}[\Sigmatb^{-1}]$ can be obtained by means of the reciprocal of the FFT
of the autocorrelation function $c(\tau) = {\rm E}\{ n[k+\tau] n[k] \}$.
Actually, in using $\Cmatb$, there is the problem that both $\ssb$ and $\xb$ are implicitly assumed to be periodic signals with period $N$. 
In general, this is
not true. Hence, boundary effects are to the expected in the computation of the FFT. However, since $\ssb$ typically has finite spatial support, 
these effects can be easily avoided by padding it with a sequence of leading zeros
longer than the correlation length of the noise. This is visible in the last three panels in
Fig.~\ref{fig:product} where the array $\Cmatb^{-1} \ssb$, that approximates the MF $\ub = \Cb^{-1} \ssb$ in Eq.~\eqref{eq:mf},
is shown when $\Cmatb$ is constructed with the correlation function shown in the first panel
and $\ssb$ is a rectangular function (shown in the same panel) that is padded with an increasing
number of leading zeros. When this number is sufficiently large, it is evident that $\Cmatb^{-1} \ssb \approx \Cb^{-1} \ssb$.
In principle, the same method could be applied to $\xb$. However, because of the noise, usually this signal has no
finite spatial support. Hence, in order to evaluate $\xb^T \Cb^{-1} \ssb$, often one wishes to compute the inverse FFT of 
${\rm DIAG}[\Sigmatb^{-1}] \odot \stb$, to remove a number of leading elements from the resulting array equal to that of the
zeros used in the padding operation, and then to calculate the scalar product with $\xb$. \\
\item In the case $\nb$ is a colored noise (i.e. $\Cb$ is not a diagonal matrix), then
the length $N$ of $\ssb$ and $\xb$ should be longer than the correlation length of $\nb$. This is clearly visible in Fig.~\ref{fig:length}
where the different performances of MF are compared when $\nb$ is the realization of a Gaussian process with a correlation length of about $100$ pixels 
and $\ssb$ is a Gaussian with $a=1$, dispersion set to three pixels that is computed, respectively, on $13$, $101$ and $301$ pixels. 
It is evident that when $N > 100$ the performance of MF becomes independent of this parameter.
The comparison is based on 
the so called {\it receiver operating characteristics} (ROC) that is a plot of $\PD$ versus $\PFA$. This kind of plot is very useful since it allows 
a direct visualization of the detection performances of a given technique. More specifically, the ROC should always be well above a $45^{\circ}$ 
straight lineline since this corresponds to a detection performance identical to that of flipping a coin, ignoring all the data.
\end{itemize}

\section{Extension to the mutiple-frequency case} \label{sec:multiple}

In the context of CMB observations, there is a further complication in that there are $M$ signals $\xb_k = \ssb_k + \nb_k$,
$\ssb_k = a_k \gb_k$, $k = 1, 2, \ldots, M$,
coming from the same sky area that are taken at different observing frequencies. Here, $a_k$ is the amplitude of the source at the
$k$th observing frequency, whereas $\gb_k$ is the corresponding spatial profile. For ease of notation, all the signals are assumed to have
the same length $N$. In general, the amplitudes $\{ a_k \} $ as well as the
profiles $\{ \gb_k \}$ are different for different $k$. However, if one sets
\begin{align}
\xb & = [\xb_1^T, \xb_2^T, \ldots, \xb_M^T]^T, \label{eq:mxb} \\
\ssb & = [\ssb_1^T, \ssb_2^T, \ldots, \ssb_M^T]^T, \label{eq:msb} \\
\nb & = [\nb_1^T, \nb_2^T, \ldots, \nb_M^T]^T, \label{eq:mnb}
\end{align}
it is possible to obtain a problem that is
formally identical to that treated in the previous section. Hence, the MF is still given by Eqs.~\eqref{eq:test1}-\eqref{eq:mf} and
is named {\it multiple matched filter} (MMF). The only difference with the classic MF is that now $\Cb$ is a $(N M) \times (N M)$ 
{\it block matrix with Toeplitz blocks} (BTB):
\begin{equation} \label{eq:covariance}
\Cb = 
\left( \begin{array}{ccc}
\Cb_{11} & \ldots & \Cb_{1M} \\
\vdots & \ddots & \vdots \\
\Cb_{M1} & \ldots & \Cb_{MM} \\
\end{array} \right),
\end{equation}
i.e. each of the $\Cb_{ij}$ blocks is constituted by a $N \times N$ Toeplitz matrix. In particular, 
$\Cb_{ii} = {\rm E}[\nb_i \nb_i^T]$ provides the autocovariance matrix of
the $i$th noise, whereas $\Cb_{ij} = {\rm E}[\nb_i \nb_j^T]$, $i \neq j$, the cross-covariance matrix between the $i$th and the $j$th ones.

In spite of these similarities, when $M > 1$ some difficulties arise. In particular, $T(\xb)$ cannot be written in a form equivalent to
Eq.~\eqref{eq:test2}. This has important consequences in the fact that if the amplitudes $\{ a_k \}$ are unknown, then $T(\xb)$
cannot be computed. In other words, if the spectral characteristics of the radiation emitted by a source are not fixed, then the MMF is not 
applicable. This forces to resort to an approach based on the maximization of the total SNR of the filtered
signals. Following this approach, model~\eqref{eq:snr} has to be modified in the form
\begin{equation} \label{eq:solu}
\ub_{\rm SNR} = \underset{ \ub }{\arg\min} [\ub^T \Cb \ub - \lambdab^T (\Sb^T \ub - \ab)],
\end{equation}
where $\ub = [\ub_1^T, \ub_2^T, \ldots, \ub_M^T]^T$ is an $(N M) \times 1$ array, 
$\lambdab = [\lambda_1, \lambda_2, \ldots, \lambda_M]^T$, $\ab = [a_1, a_2, \ldots, a_M]^T$ and $\Sb$ is a $(N M) \times M$ matrix
\begin{equation} \label{eq:S}
\Sb = \left(
\begin{array}{cccc}
\ssb_1 & \zerob & \ldots & \zerob \\
\zerob & \ssb_2 & \ddots & \zerob \\
\vdots & \ddots & \ddots & \zerob \\
\zerob & \zerob & \ldots & \ssb_M
\end{array} 
\right),
\end{equation}
with $\zerob = [0, 0, \ldots, 0]^T$ a $N \times 1$ array.
Now, since $\ab = {\rm diag}[\ab] \ooneb$ and $\Sb^T = {\rm diag}[\ab] \Gb^T$ with $\ooneb = [1, 1, \ldots, 1]^T$ 
and ${\rm diag} [\ab]$ a diagonal matrix whose diagonal
contains $\ab$, it is trivial to show that Eq.~\eqref{eq:solu} is equivalent to
\begin{equation} \label{eq:musnr}
\ub_{\rm SNR} = \underset{ \ub }{\arg\min} [\ub^T \Cb \ub - \lambdab_*^T (\Gb^T \ub - \ooneb) ],
\end{equation}
where $\lambdab_* = ({\rm diag}[\ab])^{-1} \lambdab$ and 
\begin{equation}
\Gb = \left(
\begin{array}{cccc}
\gb_1 & \zerob & \ldots & \zerob \\
\zerob & \gb_2 & \ddots & \zerob \\
\vdots & \ddots & \ddots & \zerob \\
\zerob & \zerob & \ldots & \gb_M
\end{array} 
\right).
\end{equation}
The solution is
\begin{equation} \label{eq:usnr}
\ub_{\rm SNR} = \Cb^{-1} \Gb (\Gb^T \Cb^{-1} \Gb)^{-1} \ooneb.
\end{equation}
It is evident that, contrary to MMF,  with $\ub_{\rm SNR}$ 
it is possible to obtain a statistic $T_{\rm SNR}(\xb)$,
\begin{equation}
T_{\rm SNR}(\xb) = \xb^T \ub_{\rm SNR},
\end{equation}
that is independent of the unknown source amplitude $\ab$. The price to pay is a reduced detection capability. In fact, better
results should be obtainable if the SNR of each signal was maximized. However, this is an operation that requires the knowledge of $\ab$.
For $\ub_{\rm SNR}$, it is
\begin{equation}
\PFA = Q \left( \frac{\gamma}{[ \ooneb^T (\Gb^T \Cb^{-1} \Gb)^{-1} \ooneb]^{1/2} } \right) = \alpha,
\end{equation} 
that again is a quantity independent of the source amplitude, whereas as expected the same is not true for $\PD$ 
\begin{equation}
\PD = Q \left( Q^{-1} \left( \PFA \right) - \frac{\ab^T \ooneb}
{ [\ooneb^T (\Gb^T \Cb^{-1} \Gb)^{-1} \ooneb]^{1/2} } \right).
\end{equation}

In two recent works \citet{her08a} and \citet{her08b} have proposed a ``new'' class of filters, the so called {\it matrix filters}. 
Their idea is to filter separately each of the signals $\xb_k$ in such a way to obtain unbiased estimates of $\{ a_k \}$
and at the same time to simultaneously minimize the total variance of the filtered signals. In the spatial domain, this problem can be written 
in the form
\begin{equation} \label{eq:mumfilt}
\Ub_{\rm SNR} = \underset{ \Ub }{\arg\min} \{ {\rm Tr} [\Ub^T \Cb \Ub - \Lambdab (\Gb^T \Ub - \Ib)] \},
\end{equation}
with ``${\rm Tr}$'' the {\it Trace operator},
\begin{equation}
\Ub = \left(
\begin{array}{cccc}
\ub_{11} & \ub_{12} & \ldots & \ub_{1M} \\
\ub_{21} & \ub_{22} & \ldots & \ub_{2M} \\
\vdots & \vdots & \ddots & \vdots \\
\ub_{M1} & \ub_{M2} & \ldots & \ub_{MM}
\end{array} 
\right),
\end{equation}
a $(N M) \times M$  matrix of filters and
\begin{equation}
\Lambdab = \left(
\begin{array}{cccc}
\lambda_{11} & \lambda_{12} & \ldots & \lambda_{1M} \\
\lambda_{21} & \lambda_{22} & \ldots & \lambda_{2M} \\
\vdots & \vdots & \ddots & \vdots \\
\lambda_{M1} & \lambda_{M2} & \ldots & \lambda_{MM}
\end{array} 
\right),
\end{equation}
a $M \times M$ matrix of {\it Lagrangian multipliers}.
The solution is
\begin{equation} \label{eq:umfilt}
\Ub_{\rm SNR} = \Cb^{-1} \Gb [\Gb^T \Cb^{-1} \Gb]^{-1}.
\end{equation}
Through $\Ub_{\rm SNR}$ it is possible to define a set of $M$ statistics
\begin{equation} \label{eq:uusnr}
\Tb_{\rm MatF}(\xb) = \xb^T \Ub_{\rm SNR}
\end{equation}  
that can be considered individually. However, the fact that
\begin{equation} \label{eq:uusnr1}
\ub_{\rm SNR} = \Ub_{\rm SNR} \ooneb 
\end{equation}
indicates that  $\ub_{\rm SNR}$ and $\Ub_{\rm SNR}$ essentially represent the same filter. The only difference is that, after
filtering signals $\{ \xb_k \}$, the 
latter does not compose them together as the former does. We call $\ub_{\rm SNR}$ the {\it modified multiple matched filter} (MMMF).

In appendix~\ref{sec:efficient} two procedures are presented that work in the Fourier domain and that allow a very efficient computation of
both $\ub_{\rm SNR}$ and $\Ub_{\rm SNR}$. In appendix~\ref{sec:twodimensional} the methods are extend to the two-dimensional case.

\subsection{Alternative techniques} \label{sec:alternative}

A strategy to deal with multiple-frequency images that is alternative to the approaches presented above
consists in composing signals $\{ \xb_k \}$ together in a single array $\yb$. After that,
the classic MF could be applied. Without a priori information, the most obvious method is 
\begin{equation}
\yb = \sum_{k=1}^M \Hb_k \xb_k,
\end{equation}
where $\Hb_k$ is an operator (matrix) such as $\Hb_k \gb_k = \gb$ independent of ``$k$''. In this way, the effects due to the
different instrumental beams can be avoided. As seen above, the use of $\Hb_k$ does not represent 
a problem since it does not modify the statistic $T(\xb)$. This approach, that we indicate as {\it summed-image matched filter} (SMF),
can be expected to provide results close to those of the MMF when the amplitudes
$\{ a_k \}$ as well as the level of the noises are similar. However, if this condition is not satisfied, its performances rapidly worsens
especially if
$\nb_k$ and  $\nb_l$, $k \neq l$, have some degree of correlation. This last condition is typical of CMB signals where $\nb_k = \nb_c + \eb_k$ 
with $\nb_c$ a component
independent of $k$ (i.e. the CMB contribution), and $\eb_k$ a white-noise process due to the electronic of the instrument. In this case,
it could be preferable a weighted sum as
\begin{equation} \label{eq:weighted}
\yb = \sum_{k=1}^M w_k \Hb_k \xb_k,
\end{equation} 
where
\begin{align}
\wb^T \oneb & = 0, \label{eq:constraint1} \\
\wb^T \wb & = 1, \label{eq:constraint2}
\end{align}
with $\wb=[w_1, w_2, \ldots, w_k]^T$. The first constraint implies that the contribution of $\nb_c$ in 
$\xb$ is completely removed,
whereas the second one provides a normalizing factor. We indicate this method as {\it weighted matched filter} (WMF) and
the particular case where $\wb = [\rho, \rho, \ldots, -(M-1) \rho]$
with $\rho = 1 / \sqrt{(1-M) + (1-M)^2}$, as {\it uniformly weighted matched filter} (UWMF). This last corresponds to a situation 
where only one signal is used to eliminate the shared noise component $n_c$, whereas the others are given
an identical weight. The UWMF is not very effective, but it can be useful in absence of a priori information on the spectral properties
of the sources.

\begin{figure*}
        \resizebox{\hsize}{!}{\includegraphics{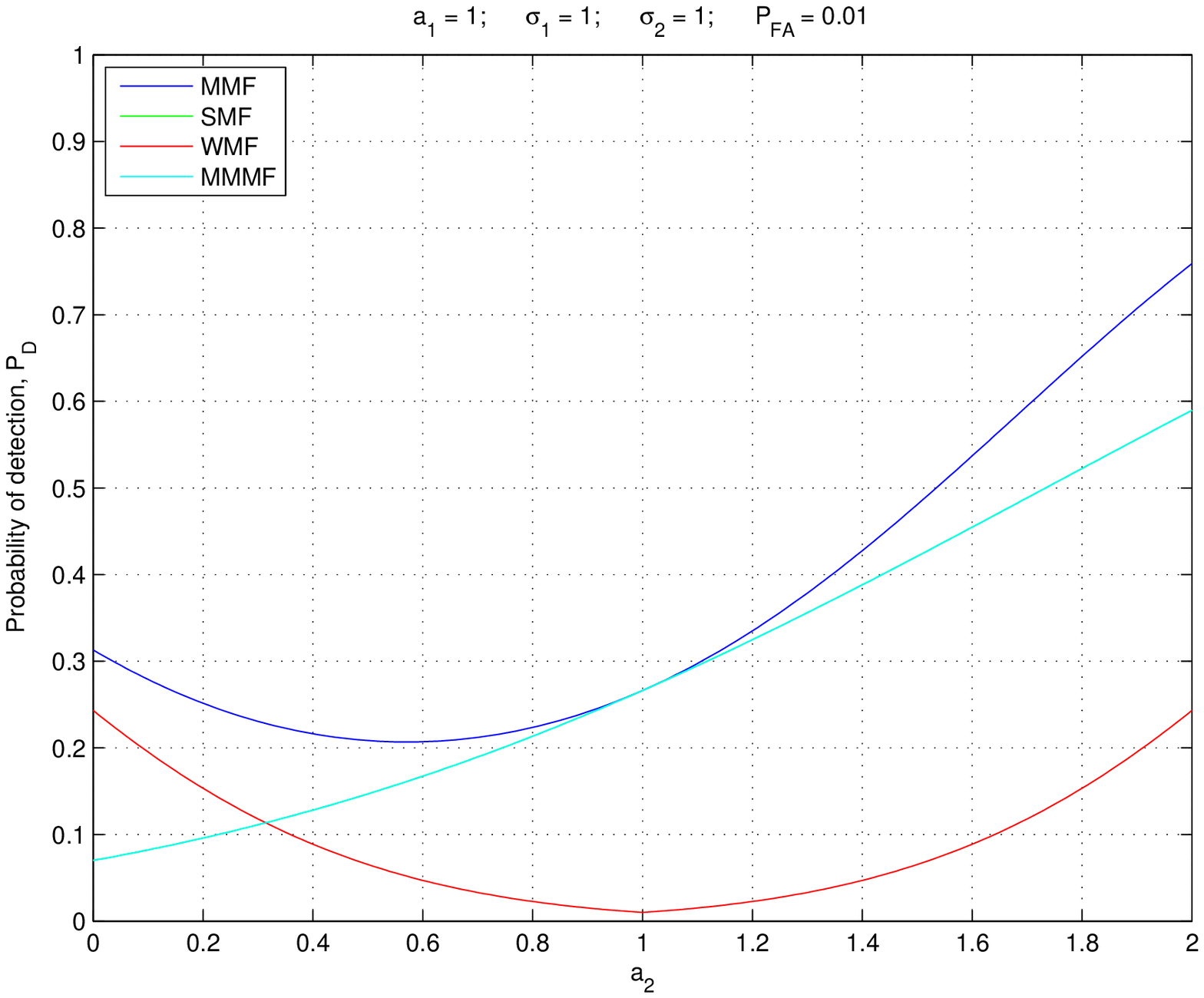}\includegraphics{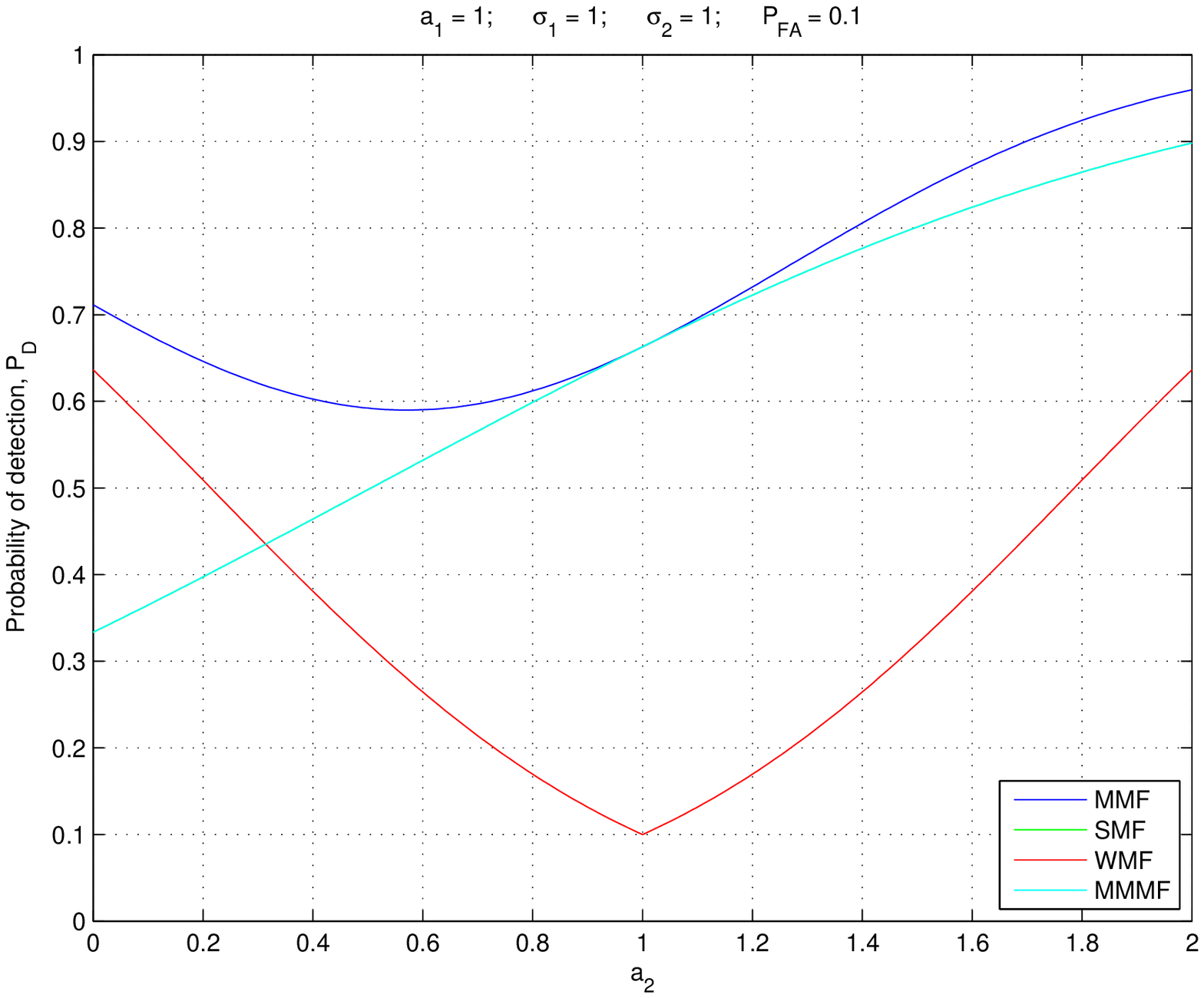}}
        \resizebox{\hsize}{!}{\includegraphics{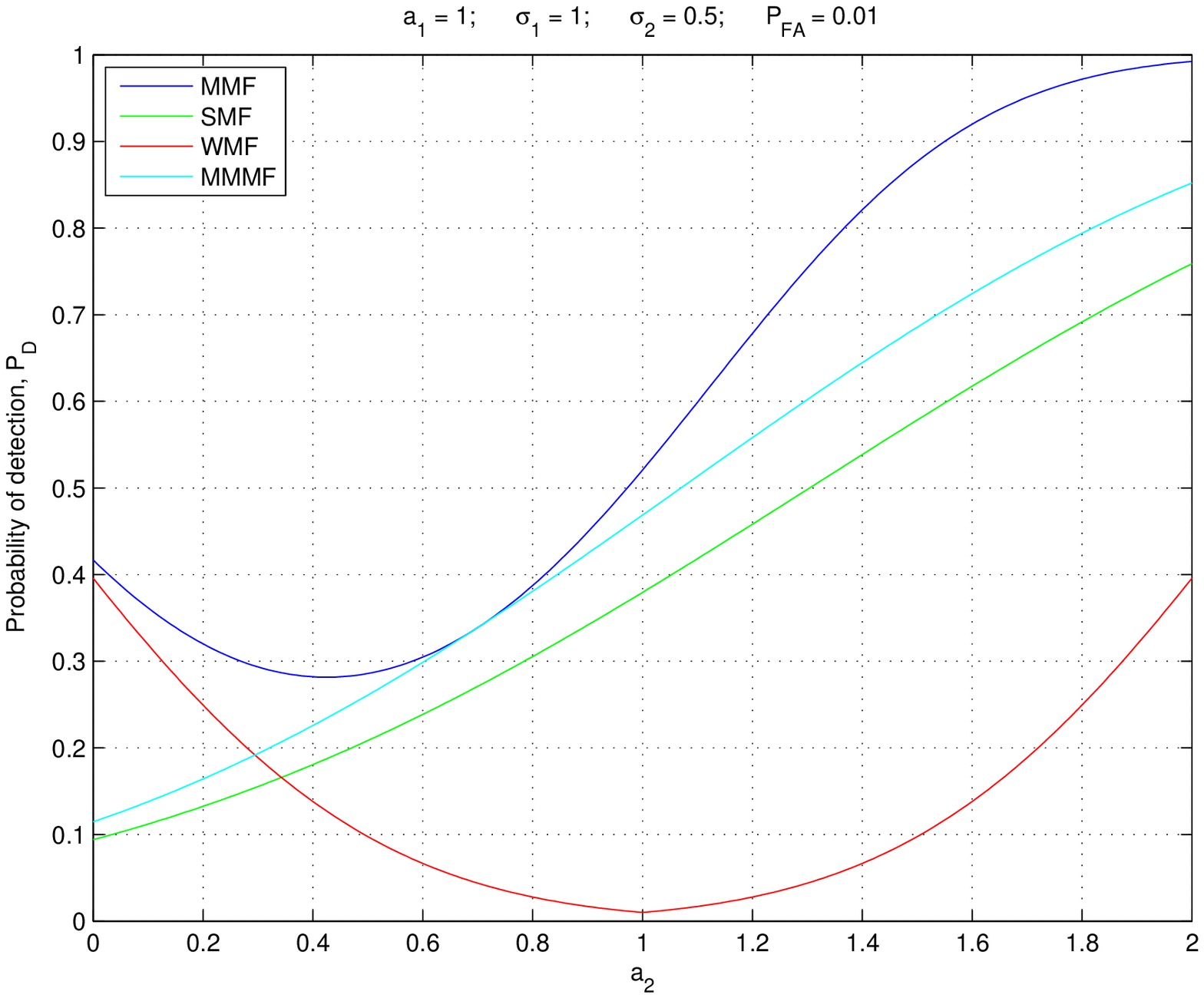}\includegraphics{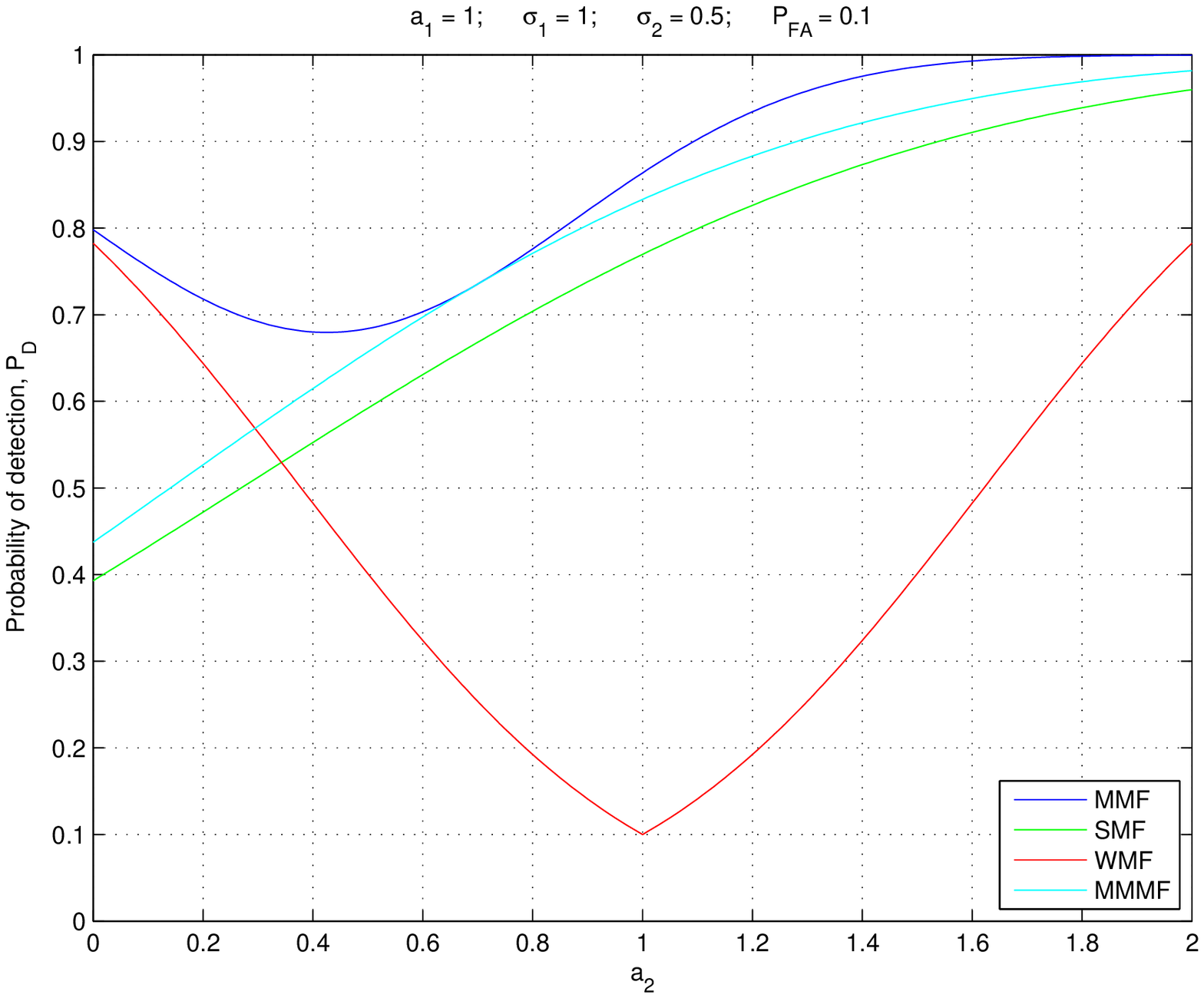}}
        \caption{Comparison of the performance of the multiple-frequency detection techniques described in Sec.~\ref{sec:multiple} in the case of
        two observing frequencies:
        {\it Probability of detection} ($\PD$) vs. the amplitude $a_2$ of the source signal at frequency $\nu_2$ when the amplitude of the
        first one at frequency $\nu_1$ is $a_1=1$.
        For the right panels the {\it probability of false alarm} ($\PFA$), i.e. the probability of a false detection, 
        is fixed to $0.01$, whereas for the left ones is fixed to $0.1$. For the top and the bottom panels, 
        the standard deviation of the noise in the two signals is set to $\sigma_1 = \sigma_2 = 1$ and $\sigma_1 = 1$ and $\sigma_2=0.5$, respectively.   
        Here, MMF = {\it multiple matched filter}, SMF = {\it summed-image matched filter},
        WMF = {\it weighted matched filter}, MMMF = {\it modified multiple matched filter}. NB. In the top panels MMMF and SMF are perfectly 
        overlapping. As expected, the superiority of MMF is unquestionable (it provides the best theoretical performance). The performance
        of the other filters depends on the relative importance of $a_2$ with respect to $a_1$.}
        \label{fig:roc001}
\end{figure*}

The performance of these methods depends critically by many factors as, for example,
the relative value of the amplitudes $\{ a_k \}$, the correlation lengths of $\{ \nb_k \}$, the relative importance of the noises $\{ \eb_k \}$ 
as well as the degree of correlation between the different 
observing frequencies. This fact is evident in Fig.~\ref{fig:roc001} that shows
the $P_D$ vs. the amplitude $a_2$ when $M=2$ and $\PFA = 0.01$ and $0.1$, respectively. Here $a_1 = 1$, $\gb$ is a Gaussian with dispersion 
set to three pixels, $\nb_k = \nb_c + \eb_k$ with $\nb_c$ a zero-mean, unit-variance Gaussian process whose autocorrelation function 
has Gaussian profile and dispersion set to ten pixels and finally $\eb_k$, $k=1,2$, two independent Gaussian white-noise processes. Two different cases
are considered for the noises $\eb_k$. In the first the noises $\eb_1$ and $\eb_2$ have the same dispersion, i.e. $\sigma_1 = \sigma_2 = 1$, whereas
in the second one $\sigma_1 = 1$ and $\sigma_2 = 0.5$. When $M=2$, the only possible weights for the WMF are either $\ub = [1/\sqrt{2}; -1/\sqrt{2}]^T$.
This is the case considered by \citet{che08}.

One indication that comes out from these examples is that, as expected, MMF outperforms all the other methods. Moreover, unless the level of
noises $ \{ e_k \}$ are similar, MMMF outperforms SMF. Heuristically, these results can be explained by the fact that through
MMF a sum of signals is computed that is weighted by means of both the intensities $\{ a_k \}$ and the noise levels $\{ \sigma_k \}$,
whereas with MMMF the weighting is based on the noise levels only. In SMF there is no weighting at all. A different situation is for WMF.
Here the two signals are subtracted. This operation provides a signal for which i) the correlated part in the noises $\nb_k$ is zeroed;
ii) the total amplitude is $a = a_1 - a_2$; iii) the variance $\sigma^2$ of the instrumental 
noise is $\sigma^2 = \sigma_1^2 + \sigma_2^2$. Since ${\rm SNR} = a^2 / \sigma^2$, it is not difficult to realize that a benefit in the 
detection capability happens when $a_2 \ll a_1$. In the case case $M > 2$, similar arguments indicate that good results can be expected when
a channel ``$l$'' is available with $a_l \ll a_{k \neq l}$. This is the case expected in CMB applications (see below).

\section{An approach for CMB experiments at high Galactic latidude} \label{sec:CMB}

In the near future some innovative ground-based experiments are planned for very high spatial resolution observations as, for example,
with the Atacama Large Millimeter/submillimeter Array  (ALMA). One important
advantage of these experiments is that, contrary to the satellite observations, they will allow a certain control of the experimental conditions. 
Hence, a full exploitation of the capabilities of this facility (and other instruments) requires a careful planning
of the observations. The recording of good quality data will allow an effective application of the chosen techniques for the detection
of point-sources. For instance, with instruments as ALMA it is possible to plan observations at high latitude fields to map sources at very high
spatial resolution in sky regions dedicated to CMB observations. In this case
a detection technique is necessary that takes into account such a specific experimental situation.

In the context of point-source detection, data can be thought as two-dimensional discrete maps $\{ \Xmatb_i \}_{i=1}^M$, each of them containing
$N_p$ pixels, corresponding to $M$ different observing frequencies (channels), with the form
\begin{equation} \label{eq:observed}
\Xmatb_i = \Smatb_i + \Nmatb_i. 
\end{equation} 
Here, $\Smatb_i$ correspond to the contribution of the point-sources at the $i$th frequency, whereas $\Nmatb_i$ denotes the corresponding
noise component. At high Galactic latitudes, the CMB component is expected to be the dominant one. Hence, $\Nmatb_i$ may be modeled with
\begin{equation} \label{eq:noise}
\Nmatb_i = \Bmatb + \Ematb_i,  
\end{equation}
where $\Ematb_i$ is the experimental noise corresponding to the $i$th channel and $\Bmatb$ the contribution of the CMB component 
that is frequency-independent. The contribution of the point-sources is assumed in the form
\begin{equation} \label{eq:point}
\Smatb_i = a_i \Gmatb,
\end{equation}
with $a_i$ the amplitude of the source to the $i$th channel. According to 
Eq.~(\ref{eq:point}), and without loss of generality, all the sources are assumed to have the same profile $\Gmatb$ independently of the 
observing frequency. In fact, although in general this will not be true, 
as written in Sect.~\ref{sec:comments}, it is possible to meet this condition 
by convolving the images with an appropriate kernel with no consequences.
In the following it is assumed that the components $\{ \Ematb_i \}$ are the realization
of stationary, zero-mean, stochastic processes.

The main feature of model~(\ref{eq:observed})-(\ref{eq:noise}) is that the CMB contribution does not change with frequency. Hence,
following the suggestion in Sec.~\ref{sec:alternative}, the WMF can be used. Of course, in order this approach be effective, it is necessary 
that the amplitude  of a given point-source is not the same in all the observing channels.

In the case  $M=2$, (i.e. only two maps are available), the only possible solution is $\wb^T = [1/\sqrt{2}; -1/\sqrt{2}]$ or
$[-1/\sqrt{2}; 1/\sqrt{2}]$. However, for $M > 2$ more degrees of freedom are available. This allows the selection of the weights in
such a way that specific conditions are satisfied. In particular, one could wish that, after the linear composition of the maps,
the quantity
\begin{equation} \label{eq:R}
R(\wb | \ab) = \frac{\ab^T \wb}{(\wb^T \Db \wb)^{1/2}},
\end{equation}
is maximized, i.e.
\begin{equation} \label{eq:problem}
\wb = \underset{ \wb }{\arg\max}  R(\wb | \ab).
\end{equation}
Here, $\Db$ is the $M \times M$ cross-covariance matrix of the noise processes whose $(i,j)$th entry $(\Db)_{ij}$ is given by
$(\Db)_{ij} = {\rm E}[{\rm VEC}^T[\Ematb_i] {\rm VEC}[\Ematb_j]]/N_p$, with 
${\rm VEC}[\Ematb]$ the operator that transforms a matrix $\Ematb$ into a column array by stacking its columns one underneath the other.
The rationale behind this choice is the
the quantity $R(\wb | \ab)$ is a measure of the amplitude of the point-source in the weighted map with respect to the standard deviation of the
measurement noise. The larger $R(\wb | \ab)$ the more prominent is the point-source with respect to the noise background; an attractive situation in
problems of source detection. The maximization, via the
{\it Lagrange multipliers} method, of $R(\wb | \ab)$ with the constraint~(\ref{eq:constraint1}) provides the
following system of non-linear equations
\begin{equation} \label{eq:solution}
(\Ib - \frac{1}{N} \oneb \oneb^T)(\ab \wb^T \Db \wb - \Db \wb \ab^T \wb) = \zerob.
\end{equation}
It can be solved through an iterative algorithm based on the Newton method
\begin{equation} \label{eq:weq}
\wb_{k+1} = \frac{\wb_k + \Deltab \wb_k}{| \wb_k + \Deltab \wb_k |} ,
\end{equation}
where
\begin{equation}
\Deltab \wb_k = -\Jb^{-1}(\wb_k) R(\wb_k).
\end{equation}
and
\begin{equation}
\Jb(\wb_k) = \sum_{i=1}^4 \Jb_i(\wb_k),
\end{equation} 
with
\begin{align}
\Jb_1(\wb) & = 2 \ab \wb^T \Db; \\
\Jb_2(\wb) & = - \ab^T \wb \Db - \Db \wb \ab^T; \\
\Jb_3(\wb) & = - \frac{2}{M} \oneb \oneb^T \ab \wb^T \Db; \\
\Jb_4(\wb) & = \frac{1}{M} (\ab^T \wb \oneb \oneb^T \Db + \oneb \oneb^T \Db \wb \ab^T).
\end{align}
Here, three points are of concern: a) the iteration can be initialized with a starting guess $\wb_0$ that contains random entries satisfying the 
constrains~(\ref{eq:constraint1})-(\ref{eq:constraint2}); b) The normalization term in Eq.~(\ref{eq:weq}) implements the 
constraint~(\ref{eq:constraint2}).
This cannot be done via a {\it Lagrange multiplier} since the array $\wb$ that maximizes the quantity $R(\wb | \ab)$ can be determined unless a 
multiplicative constant. Hence, the {\it Lagrange multiplier} corresponding to the constraint~(\ref{eq:constraint2}) should be equal to zero;
c) For the same reason, matrix $\Jb(\wb_k)$  is rank deficient, hence $\Jb^{-1}(\wb_k)$ has to be understood as
Moore-Penrose pseudo-inverse. 

Once that the map $\Ymatb$ has been produced, the 
MF can be used with no necessity to take into account the characteristics of the CMB. This is particularly useful in situations where only small
patches of sky are available and hence 
the sizes of $\Xmatb$ are much shorter than the correlation length of $\Nmatb$. As indicated earlier, we call this method the weighted matched 
filter (WMF).

As for MMF, the procedure described above needs that the array $\ab$ be specified. In fact, if the weights $\wb$ are computed 
for a source with a given $\ab$, they will be optimal only for any other source with amplitude $ \propto \ab$. However, it is the
general trend that does matter. For example, satisfactory results can be expected for the sources that present steep spectra
with similar behaviors (see below). This means that the optimization of $\wb$ can be carried out for subsets of sources for which the above 
condition approximately applies.

\subsection{Numerical experiments} \label{sec:numerical}

\begin{figure*}
        \resizebox{\hsize}{!}{\includegraphics{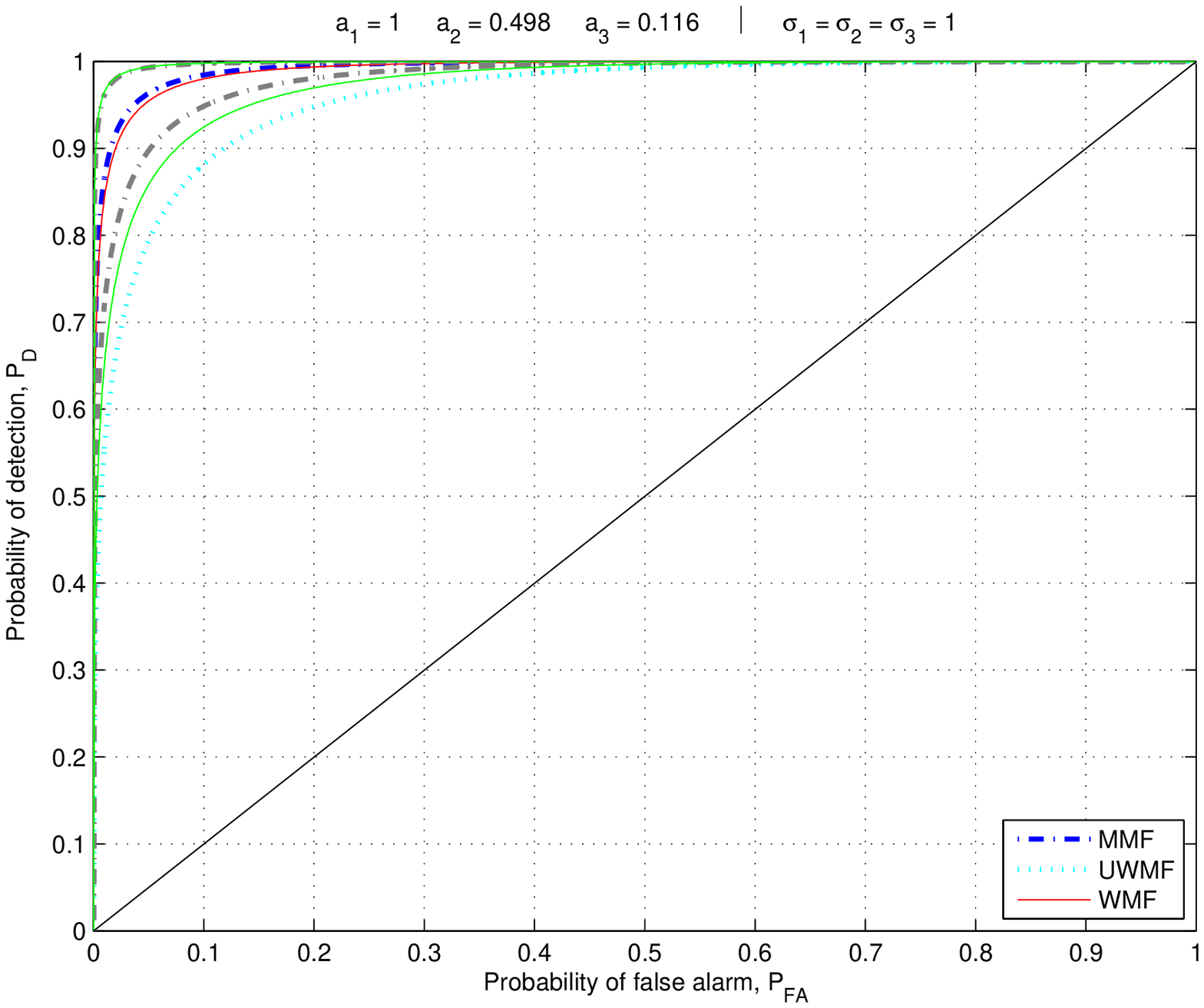}\includegraphics{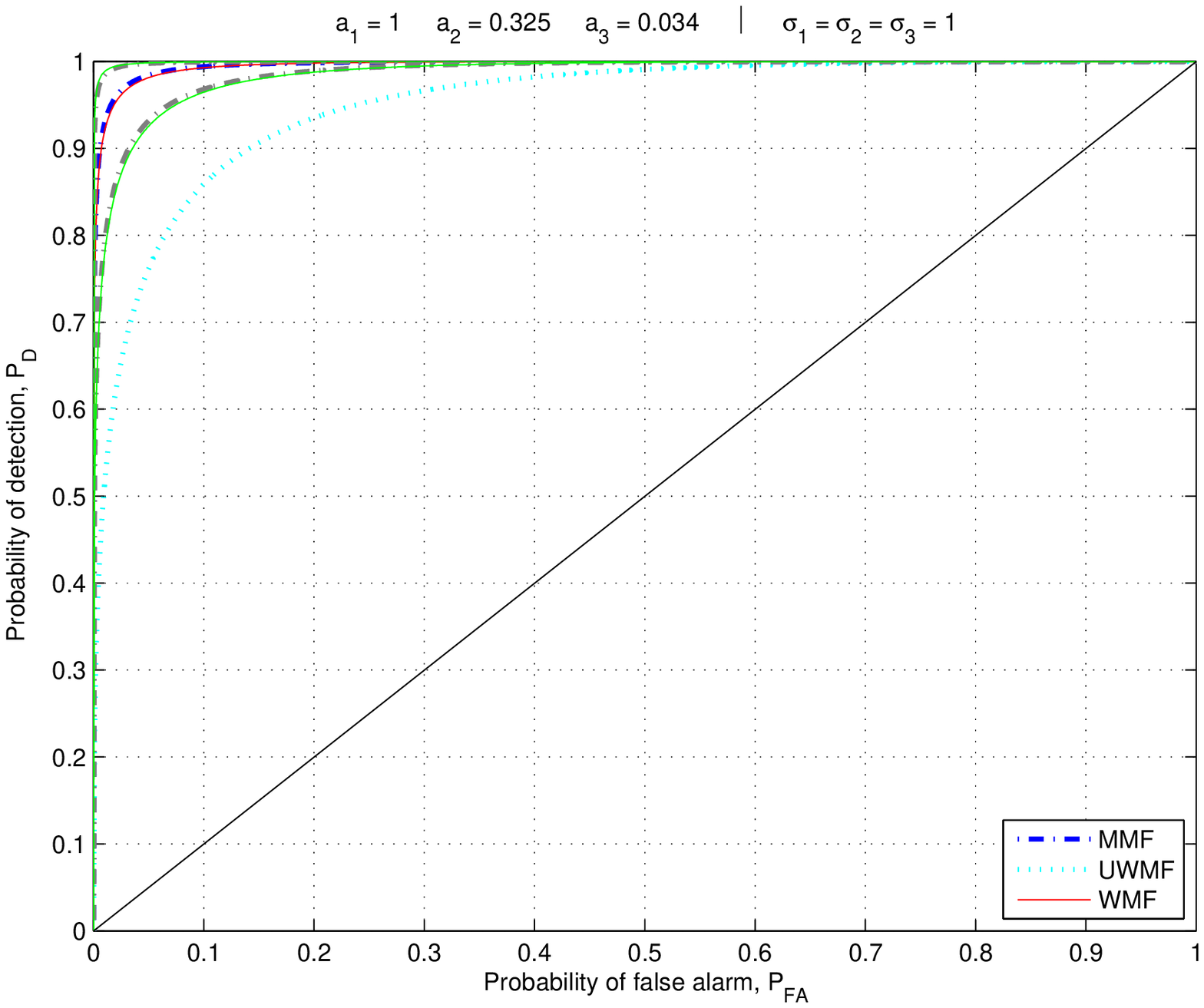}}
        \caption{Results from the numerical experiment described in Sec.~\ref{sec:numerical} relative to infrared sources (IRS) (left panel)
        and radio sources (RS) (righ panel). 
        Comparison of the ROC of the multiple matched filter (MMF) 
        with those of the weighted matched filter (WMF)
        and of the uniformly weighted matched filter (UWMF). MMF is used as benchmark since it provides the best theoretical detection performance.  
	  Here, $a_i$ and $\sigma_i$ ($i = 1 \rightarrow 30$ GHz, $i = 2 \rightarrow 70$ GHz, $i=3 \rightarrow 100$ GHz)
	  are respectively the amplitude of the source and the standard deviation of the instrumental Gaussian white-noise 
        corresponding to the $i$th observing frequency in units of the standard deviation of the CMB signal. The $90\%$ confidence envelopes are 
        shown for both MMF (gray, dot-dashed lines) and WMF (green lines) that have been obtained by computing the ROC for a set of one hundred 
        arrays $\ab + \Delta \ab $, with $\ab = [1.00; 0.50; 0.11]$ for the IRS and $\ab = [1.00; 0.33; 0.03]$ for the RS, $\Delta \ab$ a 
        Gaussian random array with 
        mean zero and covariance matrix $\Cb = 0.1 \ab^T \Ib$. For WMF the plotted ROC corresponds to the weights $\wb \approx [0.74; -0.06; -0.68]^T$
        for the IRS and $\wb = [0.78; -0.18; -0.60]^T$ for the RS. It is evident that MMF and WMF provide similar performances.}
        \label{fig:test}
\end{figure*}

In this section we present some numerical experiments to test the performances of WMF. In particular, we consider a scenario
where three different observing frequencies are available. We make the simplifying assumption that all the channels
have the same point-spread function (PSF) which is a two-dimensional circular symmetric Gaussian normalized to have a peak value equal to one and 
with a dispersion set to three pixels. In correspondence to the $i$th channel
the contribution $\Smatb_i$ due to a point-source is $\Smatb_i = a_i \Gmatb$, whereas the terms $\Cb_{ii}$ in the covariance 
matrices $\Cb$ are given by
\begin{equation}
\Cb_{ii} = \Cb_{\Bb} + \sigma_i^2 \Ib.
\end{equation}
Here $\sigma_i^2$ is the variance of the instrumental noise, assumed of Gaussian white-noise type, and $\Cb_{\Bb}$ is the covariance matrix
of the CMB component sampled with a step of $3.52'$ on a regular two-dimensional grid. 
For $i \neq j$, it is $\Cb_{ij} = \Cb_{\Bb}$. This scenarios mimics that expected for the "Low-Frequency Instrument" mounted on the 
PLANCK satellite \citep{vio03}. The available data are assumed in form of square maps containing 
$(101 \times 101)$ pixels each
\footnote{In practical CMB applications, the fact to working with square patches of sky is not a limit since, independently of the 
shape of the available maps, point-source 
detection is typically carried out on small spatial windows sliding across the sky. This is for computational reasons as well as because 
the noise contaminating the maps has no uniform spatial characteristics.}. This size
is large enough, with respect to the
correlation length of $\Cb$, to make results independent of it. For a power law spectrum as S=$a \nu^\alpha$
the amplitude of a point-source at an observed frequency 
$\nu_2$, given its amplitude at an observed frequency $\nu_1$, expressed in Thermodynamic temperature can be written as:
\begin{equation} \label{eq:a}
T_{\nu_2} = \left( \frac{\nu_2}{\nu_1} \right)^{\alpha} \frac{f(\beta_1)}{f(\beta_2)} T_{\nu_1},
\end{equation} 
where
\begin{equation}
f(\beta) = \beta^2 \frac{{\rm e}^{\beta}}{( {\rm e}^{\beta} - 1)^2},
\end{equation}
\begin{equation}
\beta = \frac{h \nu}{K T},
\end{equation}
with $h$ the Planck constant, $K$ the Boltzmann constant, $T$ the CMB temperature, and $\alpha$ is the spectral index.
\\
When expressed in {\it antenna temperature} this latter takes
the value $1.6$ for the infrared sources and $-3$ for the radio ones (dominated by synchrotron emission).
Note that this is an approximated expression, usually used in CMB experiment
to make a direct comparison between the sky temperature brightness and the sources brightness.
\\
Fig.~\ref{fig:test} compares the ROC for MMF and WMF for, respectively, the radio and the infrared sources case, 
in a situation of relatively low SNR and with the level of noise that is the same for all the channels.
The amplitudes $a_i$, computed through Eq.~\eqref{eq:a}, correspond to the $30$, $44$, $70$ GHz observing frequencies.
We allow a 10\% deviation from the amplitudes given in~\eqref{eq:a}. Hence, in the same figures
the $90\%$ confidence envelopes are shown for both MMF and WMF that have 
been obtained by computing the ROC for a set of one hundred arrays $\ab_i =  \ab + \Delta \ab_i  $, $i=1, 2, \ldots, 100$, $\Delta \ab_i$ 
a Gaussian random array with mean zero and covariance matrix $0.1 \ab^T \Ib$. For reference, the result obtainable with the UWMF is also plotted. 
The improved performances of WMF with respect to UWMF is evident. Moreover, although as expected 
the MMF is always superior to WMF, their performances are rather similar. The reason is that,
in the case of sources with steep spectra as given by Eq.~\eqref{eq:a}, there are at least two observing frequencies, say ``$l$'' and ``$k$'', 
for which $a_l \ll a_k$. As seen at the end of Sec.~\ref{sec:multiple}, this is the condition for WMF to work well. 

\section{Conclusions} \label{sec:conclusions}

In this work, the problem of point-source detection in noise background has been addressed from a theoretical perspective.
This allowed an objective comparison of the expected performance of the various techniques. From this
comparison it is evident that ``{\it in se}'' no method is superior to the other ones. The differences are due to the amount
of a priori information that they exploit. In other words, it appears that the effectiveness of a technique is not linked to its
sophistication rather to the ability in using the available a priori information. In particular, the methods based on the Neyman-Pearson criterion
can be expected to provide better performances than those based on the maximization of the signal-to-noise ratio
because they make use of the amplitude of the sources. For the same reason, an improvement in the detection capability can
be expected when more signals are available that correspond to the same sky area taking at different observing frequencies.
On the other hand, the a priori information that
at high Galactic latitudes the dominant components are the electronic noise and the CMB, with this last independent
of the observing frequency, suggests that a suboptimal approach based on an opportune linear combination of the signals
that eliminates the CMB contribution could have a performance close to MF. Hence, it is useful in practical applications
as it avoids the estimation of the covariance matrix (or the power-spectrum)
of the CMB.

\begin{figure}
        \resizebox{\hsize}{!}{\includegraphics{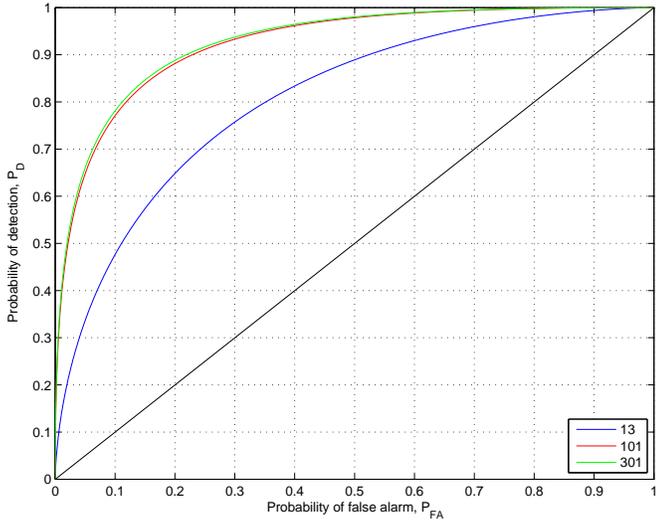}}
        \caption{Receiver operating characteristics (ROC) of Matched Filter (MF) {\it for a noise $\nb$ with a correlation
        legth of about $100$ pixels and} three different lengths of signals $\ssb$ and $\xb$ 
        (see text end of Sec.\ref{sec:comments}):
        blue, red and green lines correspond to $13$, 
        $101$ and $301$ pixels, respectively.
        The $45^{\circ}$ straight line represents a poor detection performance which is identical to 
        that of flipping a coin, ignoring all the data, i.e. better performances correspond to lines well apart from this one.
        It is evident the bad performance of MF when signals are used that are shorter
        than the correlation length of noise.}
        \label{fig:length}
\end{figure}

As last remark we would like to add the following. The superiority of a theoretical approach does not diminish the usefulness of the numerical experiments. 
It is, however, a bad habit to fix the characteristics of a statistical methodology only by means of numerical simulations.
The application of a detection technique to a set of (either real or synthetic) data requires the tuning of some parameters
that is a subjective operation. As shown in Sec.~\ref{sec:comments}, MF requires that the position
of the candidate source is known in advance, while in practical application this piece of information lacks.
The common solution consists to filter $\xb$ with $\ub$ and then to compute the statistics 
$T(\xb)$ for the peaks in the resulting signal, but, because of noise, there is no guarantee that a peak in the filtered signal marks 
the true position of a source even in the case this is effectively present. In a numerical experiment this implies 
the definition of a window, around the true position of the source, where a peak is assumed to identify a source candidate. 
The size of the window is a parameter that can have important consequences but there is not an objective criterion
to fix it. The same holds also for the other techniques. Moreover, as
remarked again in Sec.~\ref{sec:comments}, (see also Fig.~\ref{fig:length}), the performance of MF depends
on the relative size of the spatial region where the signal is sampled with respect to the correlation length of the 
noise. Hence the comparison of MF with other filters can give different results according to size of the
patch of sky that are considered. For these reasons it is risky the use of classes of filters as the mexican hat wavelet 
family \citep{lop06a,lop06b} that lack any theoretical justification and 
whose effectiveness is supported {\it only} by numerical experiments.

\clearpage
\begin{figure*}
        \resizebox{\hsize}{!}{\includegraphics{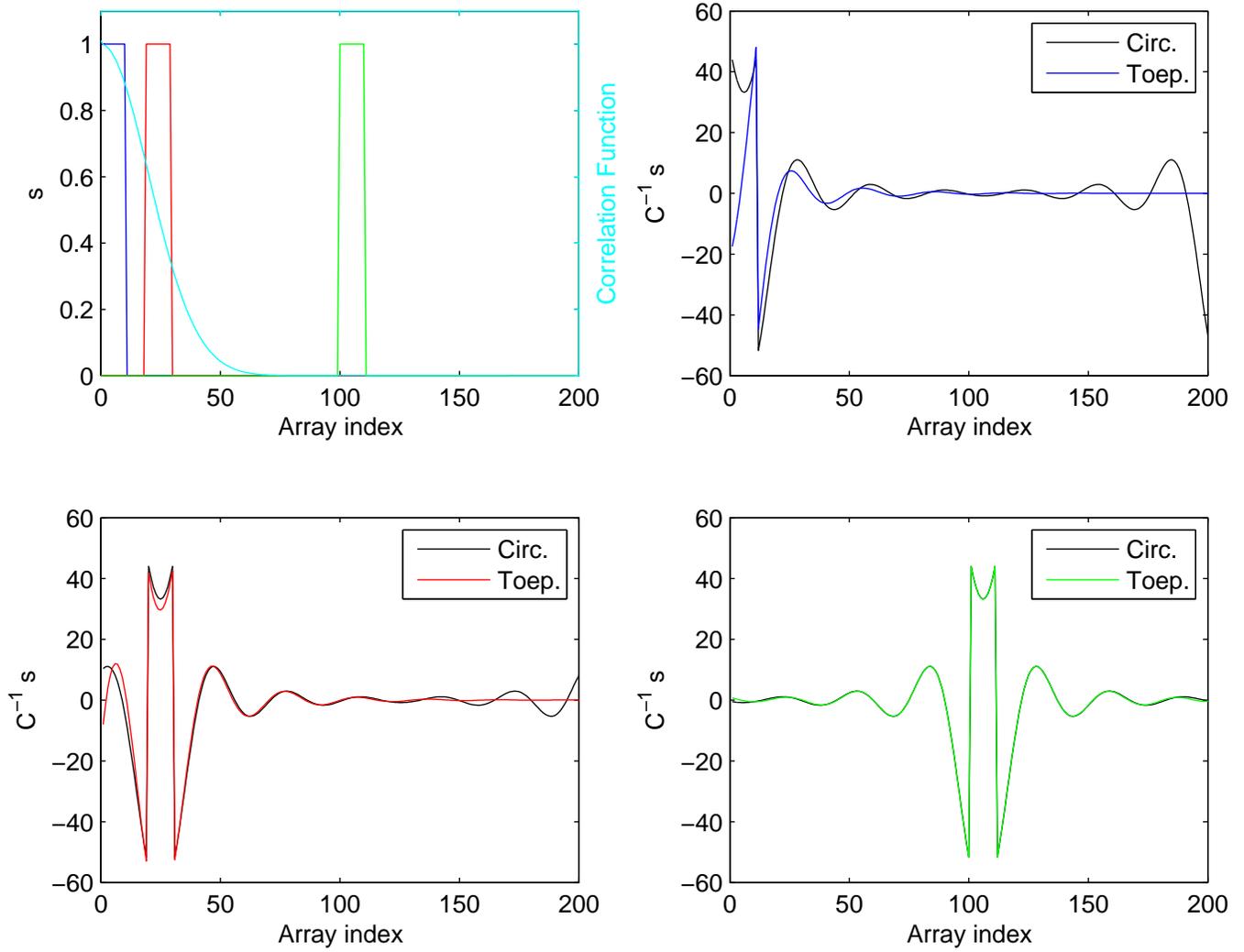}}
        \caption{Experiment that shows how the matched filter $\ub = \Cb^{-1} \ssb$, Eq.~\eqref{eq:mf}, with $\Cb$ the 
        covariance matrix defined in Eq.~\eqref{eq:C} (a Toeplitz matrix), can be well approximated using a circulant matrix $\Cmatb$
        if signal $\ssb$ is padded with a number of leading zeros sufficiently large (see Sec.~\ref{sec:comments}). Top-left panel: signal $\ssb$ 
        (a rectangular function) padded with $0$ (blue line), $20$ (red line) and $100$ (green line) leading zeros. 
        For reference,
        the covariance function (cyan line) is shown that is used to form the Toeplitz matrix $\Cb$ and its {\it circulant} approximation $\Cmatb$.
        The top-right, bottom-left and bottom-right panels compare array $\Cb^{-1} \ssb$ with $\Cmatb^{-1} \ssb$ for the three zero-padding 
        situations mentioned above (respectively, $0$, $20$ and $100$ zeros). It is evident that when the number of zeros is larger than
        the correlation length of the noise $\nb$ then $\Cb^{-1} \ssb$ and $\Cmatb^{-1} \ssb$ are almost indistinguishable.}
        \label{fig:product}
\end{figure*}
\clearpage

\clearpage

\clearpage

\clearpage
\appendix
\section{Efficient numerical implementations of the MMMF: one dimensional case} \label{sec:efficient}

Filters $\ub_{\rm SNR}(\xb)$ and $\Ub_{\rm SNR}(\xb)$ given, respectively, by Eqs.~\eqref{eq:usnr} and
Eq.\eqref{eq:umfilt} can be efficiently computed in the Fourier domain. If in the BTB matrix
$\Cb$ given by Eq.~\eqref{eq:covariance} the Toeplitz blocks
are approximated with circulant matrices $\Cmatb_{ij}$, then each of them can be diagonalized by means of
the $(N M) \times (N M)$ matrix
\begin{equation}
\Fmatfb = \Ib_{(M)} \otimes  \Fb_{(N)},
\end{equation} 
with ``$\otimes$'' the Kronecker product and $\Ib_{(M)}$ the $M \times M$ identity matrix.
In this way a matrix
\begin{equation}
\Sigmatb = 
\left( \begin{array}{ccc}
\Sigmatb_{11} & \ldots & \Sigmatb_{1M} \\
\vdots & \ddots & \vdots \\
\Sigmatb_{N1} & \ldots & \Sigmatb_{MM} \\
\end{array} \right)
\end{equation}
is obtained where $\Sigmab_{ij}$ are diagonal blocks containing the eigenvalues of $\Cmatb_{ij}$.
The quantities $\Gb^T \Cmatb^{-1}$ and $\Gb^T \Cmatb^{-1} \Gb$,
can be computed firstly by solving the system of linear equations $\Sigmatb \Ztb = \Gtb$, 
and then through the product $\Gtb^H \Ztb$ (or through the procedure described in Sec.~\ref{sec:comments}). 
Since matrix $\Sigmatb$ is highly structured and sparse, the computational load is not excessive. 
This approach is suited to be implemented in high-level programming languages as MATLAB that allow a friendly handling of
arrays and matrices. 

A more efficient algorithm, but suited for low-level programming languages as C or FORTRAN,
is obtainable rearranging the elements of $\xb$, $\ssb$, $\nb$ and $\ub$ according to the so called {\it row rollout} order, i.e.,
\begin{equation}
\xbl = [\xbl[0], \xbl[1], \ldots, \xbl[N-1]]^T,
\end{equation}
with $\xbl[i] = [x_1[i], x_2[i], \ldots, x_M[i]]^T$, and similarly for $\ssbl$, $\nbl$ and $\ubl$. After that, 
models.~\eqref{eq:musnr} and \eqref{eq:mumfilt}, as well as the corresponding solutions~\eqref{eq:usnr} and~\eqref{eq:umfilt},
still holds with $\Cb$ and $\Gb$  replaced, respectively, by
\begin{equation} \label{eq:Cs}
\Cbl = \left(
\begin{array}{cccc}
\Cbl[0] & \Cbl[-1] & \ldots & \Cbl[-(N-1)] \\
\Cbl[1] & \Cbl[0] & \ldots & \Cbl[-(N-2)] \\
\vdots & \vdots & \ddots & \vdots \\
\Cbl[N-1] & \Cbl[N-2] & \ldots & \Cbl[0]
\end{array}
\right),
\end{equation}
where $\Cbl[\tau]={\rm E}[\nbl[i] \nbl^T[i+\tau]]$, and
\begin{equation} \label{eq:Ss}
\Gbl = [{\rm DCS}_0[\gbl_1], {\rm DCS}_1[\gbl_2], \ldots, {\rm DCS}_{M-1}[\gbl_M]],
\end{equation}
with
\begin{equation}
\gbl_k = [\gl_k[0], \zerob^T_{(M-1)}, \gl_k[1], \zerob^T_{(M-1)}, \ldots, \gl_k[N-1], \zerob^T_{(M-1)}]^T. 
\end{equation}
Here, ${\rm DCS}_l[.]$ denotes the {\it down circulant shifting} operator that circularly down shifts the elements of a column array 
by $l$ positions. Now, it can be shown again \citep[see][ page 504]{kay98} that, if one sets
\begin{equation}
\Fmatfb = \Fb_{(N)} \otimes \Ib_{(M)},
\end{equation}
then
\begin{equation} \label{eq:VCV}
\Fmatfb \Cbl \Fmatfb^H \approx \Sigmatbl,
\end{equation}
where $\Sigmatbl$ is a block diagonal matrix
\begin{equation}
\Sigmatbl = \left( \begin{array}{ccl}
\Sigmatbl_0 & &  \Large{\zerob} \\
& \ddots & \\
\Large{\zerob} & & \Sigmatbl_{N-1}
\end{array} \right),
\end{equation}
with
\begin{equation}
\Sigmatbl_i = \left( \begin{array}{cccc}
\Pb_{11}(f_i) & \Pb_{12}(f_i) & \ldots & \Pb_{1M}(f_i) \\
\vdots & \vdots & \ddots & \vdots \\
\Pb_{M1}(f_i) & \Pb_{M2}(f_i) & \ldots & \Pb_{MM}(f_i)
\end{array} \right),
\end{equation}
$f_i = i/N$, $i=0, 1, \ldots, N-1$.
Here, $\Pb_{kl}(f_i)$ represents the cross-power-spectrum at frequency $f_i$ between $\nb_k$ and $\nb_l$.
Similarly, for a given signal $\rbl$, the array $\Fmatfb^H \rbl$ provides the corresponding FFT,
\begin{equation} \label{eq:Vr}
\Fmatfb \rbl = \rtbl = \left(
\begin{array}{c}
\rtbl[f_0] \\
\rtbl[f_1] \\
\vdots \\
\rtbl[f_{N-1}] \\
\end{array}
\right).
\end{equation}
From these considerations, it results that
\begin{equation} \label{eq:Utbl}
\Utbl_{\rm SNR} = \Sigmatbl^{-1} \Gtbl (\Gtbl^H \Sigmatbl^{-1} \Gtbl)^{-1},
\end{equation}
with
\begin{equation}
\Gtbl = [{\rm DCS}_0[\gtbl_1], {\rm DCS}_1[\gtbl_2], \ldots, {\rm DCS}_{M-1}[\gtbl_M]].
\end{equation}
Here, the advantage is represented by the fact that
\begin{equation}
\Sigmatbl^{-1} = \left( \begin{array}{ccl}
\Sigmatbl^{-1}_0 & &  \Large{\zerob} \\
& \ddots & \\
\Large{\zerob} & & \Sigmatbl^{-1}_{N-1}
\end{array} \right),
\end{equation}
i.e., the inversion of $\Sigmatbl$, that is a large matrix with size $(N M) \times (N M)$, can be obtained through the inversion
of a number $N$ of much smaller $M \times M$ blocks. This fact, coupled with the structure of $\Gtbl$,
allows to compute efficiently $\Utbl_{\rm SNR}$ through block-matrix operations.
Equation~\eqref{eq:Utbl} provides the discrete version of the result by \citet{her08a} and \citet{her08b}. 
Finally, as for the spatial domain, it is
\begin{equation} \label{eq:utbl}
\utbl_{\rm SNR} = \Utbl_{\rm SNR} \oneb.
\end{equation}

\section{Extension of MF and MMMF to the two-dimensional case} \label{sec:twodimensional}

\subsection{Single-frequency observations}

The extension of MF to the two-dimensional signals $\Xmatb$ and $\Smatb$ is conceptually trivial. If one sets
\begin{align}
\ssb & = {\rm VEC}[\Smatb]; \label{eq:stack1} \\
\xb & = {\rm VEC}[\Xmatb]; \label{eq:stack2} \\
\nb & = {\rm VEC}[\Nmatb], \label{eq:stack3}
\end{align}
formally the same problem is obtained as that given by Eq.~\eqref{eq:decision}. However, again some computational issues come out. 
The question is that, even for moderately sized signals, matrix $\Cb$
becomes rapidly huge. In fact, if $\Smatb$ contains $N_p$ pixels, then $\Cb$ is a $N_p \times N_p$ matrix. Similarly to the
one-dimensional case, the computational burden can be alleviated by resorting to the Fourier domain. If $\Smatb$ is 
a $N_r \times N_c$ rectangular map, then $\Cb$ is a matrix of type
{\it block Toeplitz with Toeplitz blocks} (BTTB), and it can approximated with a matrix of type {\it block circulant with circulant blocks} (BCCB).
In fact, a BCCB matrix $\Cmatb$ can be diagonalized through
\begin{equation}
\Fmatb \Cmatb \Fmatb^H = \Sigmatb,
\end{equation}
where
$\Fmatb = \Fb_{N_r} \otimes \Fb_{N_c}$, and $\Sigmatb$ is again a diagonal matrices containing the eigenvalues
of $\Cmatb$. 
The good news is that these eigenvalues can be obtained through the application of the  ${\rm VEC}[.]$ operator
to the two-dimensional FFT of the autocovariance function $c(\tau_1, \tau_2) = {\rm E}\{ \Nmatc(j+\tau_1, l+\tau_2) \Nmatc(j,l) \}$ of $\Nmatb$. 
Since also $ \Fmatb$ is a unitary matrix, it is
\begin{align}
\xb^T \Cb^{-1} \ssb & \approx (\xb^T \Fmatb^H) (\Fmatb \Cmatb \Fmatb^H)^{-1} (\Fmatb \ssb) = \\
& = \xtb^H \Sigmatb^{-1} \stb = \xtb^H ({\rm DIAG}[\Sigmatb^{-1}] \odot \stb). 
\end{align}
Here, symbol ``$\widetilde{~~}$'' now indicates the two-dimensional FFT.
Similar problems and solutions as in Sec.~\ref{sec:comments} hold concerning the fact that, with the use of $\Cmatb$, both $\Smatb$ and 
$\Xmatb$ are implicitly assumed to be periodic functions along each dimension with period $N_r$ and $N_c$, respectively. 
Once computed, filter $\ub = \Cb^{-1} \ssb$, that is a $(N_r N_c) \times 1$ array, can be converted in its original two-dimensional form
simply columwise reordering its elements in a $N_r \times N_c$ matrix $\Umatb$ (i.e., the inverse operation of the ${\rm VEC}[.]$ operator).

\subsection{Multiple-frequency observations}

In the case of multi-frequency observations, the situation becomes even worst 
since MF has to be applied to $M$ signals at the same time. Again, through
\begin{align}
\ssb & = {\rm VEC}\left[ {\rm VEC}[ \Smatb_1 ], {\rm VEC}[ \Smatb_2 ], \ldots, {\rm VEC}[ \Smatb_M ] \right]; \label{eq:stack12} \\
\xb & = {\rm VEC}\left[ {\rm VEC}[ \Xmatb_1 ], {\rm VEC}[ \Xmatb_2 ], \ldots, {\rm VEC}[ \Xmatb_M ] \right]; \label{eq:stack22} \\
\nb & = {\rm VEC}\left[ {\rm VEC}[ \Nmatb_1 ], {\rm VEC}[ \Nmatb_2 ], \ldots, {\rm VEC}[ \Nmatb_M ] \right], \label{eq:stack32}
\end{align}
it is possible to obtain a problem that is
formally identical to that given by Eq.~\eqref{eq:decision}. The only difference is that now $\Cb$ in Eq.~\eqref{eq:covariance} 
$\Cb$  is a $(M N_p) \times (M N_p)$ 
block matrix. In the case signals $\{ \Smatb_i \}$ are two-dimensional $N_r \times N_c$ maps, then  
each of the $\Cb_{ij}$ blocks is constituted by a $(N_r N_c) \times (N_r N_c)$  BTTC matrix. In particular, 
$\Cb_{ii}$ provides the autocovariance matrix of
the $i$th image, whereas $\Cb_{ij}$, $i \neq j$, the cross-covariance matrix between the $i$th and the $j$th ones. If, again, each
BTTB block $\Cb_{ij}$ is approximated with a BCCB matrix $\Cmatb_{ij}$, then matrix
\begin{equation} \label{eq:FF}
\Fmatfb = \Ib_{(M)} \otimes \Fmatb, 
\end{equation} 
can be used to diagonalize each of the blocks $\Cmatb_{ij}$ in $\Cb$. In this way, the statistics $T(\xb) = \xb^T \Cmatb^{-1} \ssb$,
can be computed firstly by solving the system of linear equations $\Sigmatb \ztb = \stb$, 
\begin{equation}
\Sigmatb = 
\left( \begin{array}{ccc}
\Sigmatb_{11} & \ldots & \Sigmatb_{1M} \\
\vdots & \ddots & \vdots \\
\Sigmatb_{N1} & \ldots & \Sigmatb_{MM} \\
\end{array} \right),
\end{equation}
that is highly structured and sparse 
(i.e., not computational demanding), and then through $\xtb^H \ztb$. 
Alternatively, if $\xbl$, $\ssbl$ and $\nbl$ indicate the arrays~\eqref{eq:stack12}-\eqref{eq:stack32} with the elements that are 
rearranged in {\it row rollout} 
order, and $\Cbl$ is the covariance function of $\nbl$, then $T(\xbl) = \xtbl^H \Sigmatbl^{-1} \stbl$, $\Sigmatbl \approx \Fmatfb \Cbl \Fmatfb^H$,
can be efficiently computed through block-matrix operations exploiting the fact that $\Sigmatbl$ is a block diagonal matrix.

Both approaches can be used to compute filters $\ub_{\rm SNR}(\xb)$ and $\Ub_{\rm SNR}(\xb)$. In particular,
if the elements of $\xbl$, $\ssbl$ and $\nbl$ are arranged in {\it row rollout} order, a solution formally identical to Eq.~\eqref{eq:Utbl}
can be obtained if matrix
\begin{equation}
\Fmatfb = \Fmatb \otimes \Ib_{(M)}
\end{equation}
is used in Eqs.~\eqref{eq:VCV} and \eqref{eq:Vr}. 
Again, this result coincides with that provided by \citet{her08a} and \citet{her08b}. 

\end{document}